# Prediction of novel two-dimensional rare-earth material with room-temperature ferromagnetism and large perpendicular magnetic anisotropy

Haoyi Tan, [a] Guangcun Shan, [a,*] Jiliang Zhang[b,c]

[a] School of Instrumentation Science and Opto-electronics Engineering & Institute of Precision Instrument and Quantum Sensing, Beihang University, Beijing 100191, China.

[b] Department of Materials Science and Engineering, City University of Hong Kong, Kowloon Tong, Hong Kong SAR.

[c] Department of Materials Science and Engineering, Dalian Jiaotong University, Dalian, China

**Abstract**

Novel 2D ferromagnets with high Curie temperature and large perpendicular magnetic anisotropy are especially attractive owing to the future promising application in modern spintronics, but meanwhile the 2D ferromagnetic materials with high Curie temperature and large perpendicular magnetic anisotropy are rarely reported. Based on density functional theory (DFT) calculations, we predict a new kind of 2D ferromagnetic materials - $GdB_2N_2$, which possesses large magnetic moment (~7.87 μB/f.u.), high Curie temperature (~335 K) and large perpendicular magnetic anisotropy (~10.38 meV/f.u.). Biaxial strain ranging from -0.5% to 5% and different concentrations of charge-carrier doping (⩽0.5 e/h per f.u.) are applied to reveal the influence on the Curie temperature and magnetic anisotropy energy (MAE). Besides, magnetic coupling process within $GdB_2N_2$ is found to be via a Ruderman-Kittel-Kasuya-Yosida (RKKY) mechanism. In summary, our work here predicts a novel 2D rare-earth material GdB$_2$N$_2$, which not only enriches the category of 2D room-temperature ferromagnets, but also proposes a new possibility of combining traditional 2D materials and rare-earth materials to achieve more intriguing magnetic properties, finally it carves out the path for the next-generation spintronic devices and sensors.

1. Introduction

In recent years, two-dimensional (2D) materials represented by graphene have drawn much interest because of their unique structures and interesting physical properties [1,2]. The reduced dimensionality provides 2D materials tiny thickness and tunable fascinating properties, compared to their traditional 3D counterparts. Up to now, dozens of 2D materials have been successfully synthesized and practically studied, including *h*-BN [3], transition metal dichalcogenides (TMDCs) and MXenes [4,5], etc. Meanwhile, 2D materials with ferromagnetism have been intensively studied and discovered one after another [6], which includes CrI$_3$ [7], Cr$_2$Ge$_2$Te$_6$ [8], VSe$_2$ [9], Fe$_3$GeTe$_2$ [10], hematene Fe$_2$O$_3$ [11], chromene Cr$_2$O$_3$ and CrBr$_3$ [12,13]. Ferromagnetic 2D materials are regarded as promising platform for high-performance spintronic devices and sensors. Furthermore, van der Waals heterojunctions by stacking different 2D materials have also greatly broaden the magnetic applications [14].

Rare-earth elements, especially gadolinium and europium with half-filled and well-localized 4f subshell, usually own high saturation magnetism and strong magnetocrystalline anisotropy [6]. Therefore, it is meaningful to design 2D rare-earth magnets with high Curie Temperature. Up to now, only few rare-earth 2D ferromagnetic materials or heterojunctions have been reported [15-24]. Last year, we have recently predicted the novel EuBr/graphene heterojunctions with large intrinsic-ferromagnetism of nearly 7.0 μ$_B$ per Eu and perpendicular magnetic anisotropy, yet both the Curie temperature and thermal stability of EuBr/graphene heterojunctions have been reported to be much below the expectation of room-temperature [24]. In 2020, Wang et al. predicted a GdI$_2$ monolayer with a large magnetization (~8 μB/f.u.), high Curie temperature (~241 K) and large MAE (~553 μeV/Gd) [22]. It was revealed that the combination of the strong Gd$_{4f}$-Gd$_{5d}$ and Gd$_{5d}$-I$_{5p}$-Gd$_{5d}$ interactions leads to strong ferromagnetism in the 2D limit. Subsequently, Liu et al. predicted the monolayer GdCl$_2$ and GdBr$_2$ with the high Curie temperature of 224 K and 229 K, respectively [23]. Nevertheless, the Curie temperatures of the monolayer GdI$_2$, GdCl$_2$ and GdBr$_2$ above are still below room-temperature, and the easy axial of GdCl$_2$ and GdBr$_2$ is along the in-plane direction, all of which restricts their future practical applications for magnetic recording.

Above all, the 2D ferromagnets with huge saturation magnetism, high Curie temperature above room-temperature and large perpendicular magnetic anisotropy are still very rare. In this study, based on DFT calculations, we predict a novel 2D GdB$_2$N$_2$ material, which simultaneously owns huge saturation magnetism (~7.87 μ$_B$/f.u.), high Curie temperature (~335 K), large MAE (~10.38 meV/f.u.) and extraordinary stability. Our theoretical work provides the new perspective of 4f magnets by combining rare-earth elements and traditional 2D materials, which opens up the new way of next-generation spintronic devices and sensors.

## 2. Methods

We performed the first-principles calculations based on DFT with the Perdew-Burke-Ernzerhof (PBE) functional in the generalized gradient approximation (GGA) using the Vienna ab initio simulation package (VASP) [25-28]. The standard pseudopotential was used with the valence electron configurations. It is noteworthy that due to the half-filled 4f orbitals - 4f$^7$ in Gd, the first-principles calculations are still well efficient to describe and predict the electronic and magnetic properties of GdB$_2$N$_2$. The cut-off energy for the plane wave basis was set as

550 eV and a 17×17×1 Gamma centered k-mesh was used to sample the first Brillouin zone. The accuracy of electronic self-consistency was set to be $10^{-6}$ eV between the two electronic steps. Because of the strongly corrected electrons, a Hubbard on-site Coulomb potential was set for Gd to U-J = 8.0 eV, as referred and suggested according to the previous work [22,23,29,30]. The structures were fully relaxed until the forces on each atom converged to 0.001 eV/Å. The spin-orbit coupling (SOC) effect was taken into account during the calculations of MAE and band structure due to the Gd as heavy element. A vacuum layer of more than 15 Å is added to shield the interactions between periodic neighboring layers. Additionally, DFT-D3 corrections were included to consider the interlayer van der Waals interactions [31,32].

## 3. Results and discussion

The 2D $GdB_2N_2$ possesses the hexagonal crystal structure with space group of P-6m2. As shown in figure 1(a), the $GdB_2N_2$ consists of three in-plane atomic layers. The two BN layers are mirror symmetrical to the middle Gd layer, and the Gd ions are in the middle of the upper and lower N ions. The lattice parameter is 0.266 nm, which is slightly longer than that of *h*-BN crystals (~0.250 nm) [33]. The distance between the Gd layer and BN layer is 0.260 nm, and the bond length of B-N is 0.154 nm. The distance between the neighboring Gd ions is 0.266 nm, indicating the direct exchange interactions are weak. By comparing the energy of $GdB_2N_2$ with different magnetic configurations, including ferromagnetic (FM), ferrimagnetic (FIM) and two antiferromagnetic states (AFM1 and AFM2), FM configuration is finally identified as the energy-optimal state with the energy difference of 198.05 meV/Gd. Different magnetic configurations are shown in figure 1(b) and the calculated corresponding total energy is listed in table S1. The energy of the primitive cells with magnetic moment along different directions is also calculated and listed in table S2. The [001] direction with the lowest energy corresponds to the easy magnetization axis of GdB$_2$N$_2$, with the MAE of 10.38 meV per Gd, much larger than that of the most studied 2D magnets, such as CrI$_3$ (~0.70 meV per Cr) [34] and Fe$_3$GeTe$_2$ (~0.67 meV per Fe) [10]. The orbital resolved MAE of GdB$_2$N$_2$ is calculated and shown in figure S1. The Gd-p and Gd-f orbitals provide the main positive MAE, while Gd-d orbitals provide negative MAE. The contributions of B-p and N-p orbitals to MAE are little.

For identifying the type of interactions of atoms within GdB$_2$N$_2$, the electron localization functional (ELF) with the value ranging from 0 to 0.7, is plotted in figure 1(c) to quantitatively indicate different localizations of electrons. It can be found that the electrons highly localize between the B and N ions with the value of ELF above 0.7, which means the interactions are strong covalent bonds. Comparatively, the electrons between the Gd ions tend to delocalize, which indicate the relatively weak interactions. But interestingly, there are even some electrons localizing between the Gd and BN layers, which indicates that a kind of complex interlayer coupling exists within heterojunctions. For further discussion, electron density difference of the whole GdB$_2$N$_2$, Gd and BN layer are separately analyzed as shown in figure 2(a). In addition, Bader charge analysis is also applied to quantitatively analyze electron transfer: 0.42$e$ of each Gd transfers to N ions, and 1.60$e$ of each B transfer to N ions. That is to say, each N obtains 1.82$e$ in GdB$_2$N$_2$.

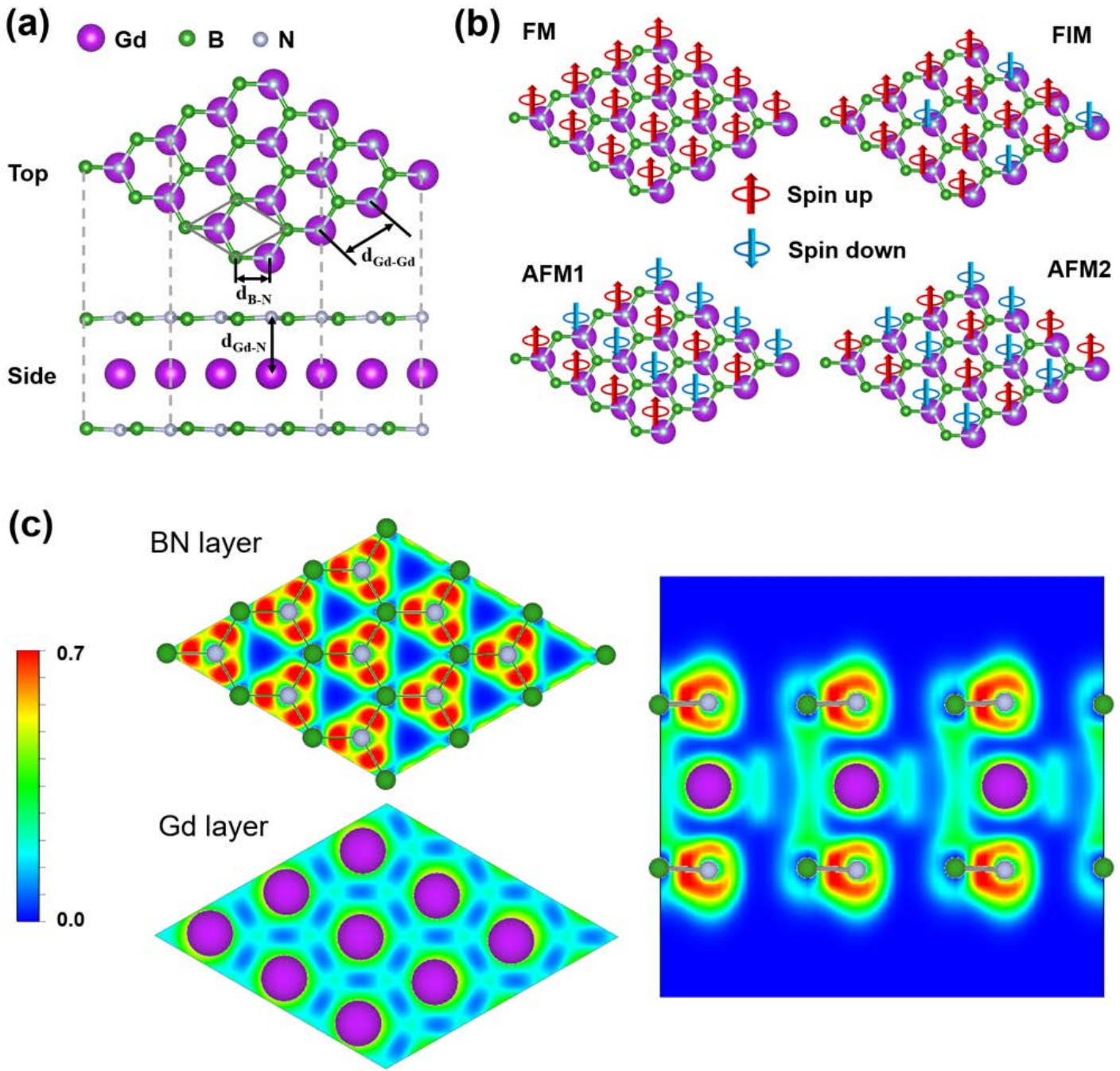

**Figure 1 (a)Geometric structures, (b)different magnetic configurations and (c) ELF of BN layer, Gd layer and cross section.**

Next, we evaluate the stability of newly predicted $GdB_2N_2$, which is essential for experimental synthesis and practical applications in future. Firstly, the cohesive energy of $GdB_2N_2$ is calculated to be 6.06 eV/atom, which is larger than that of other table 2D materials, such as silicene (~5.16 eV/atom) and germanene (~4.15 eV/atom) [35]. The large cohesive energy confirms the thermodynamic stability. Secondly, phonon spectrum is plotted in figure

2(b). That no imaginary frequency appears in graph means the GdB$_2$N$_2$ is kinetically stable [36]. AIMD simulation is also applied based on the canonical ensemble (NVT) with a Nosé thermostat for temperature control as shown in figure 2(c) [37], 5×5 supercells of GdB$_2$N$_2$ are put at 500 K with a total of 10 ps at 1 fs per step. During the whole simulations, the range of energy fluctuation is small (< 0.1 eV/atom) and the final structure is also well preserved. The above results manifest the GdB$_2$N$_2$ has good thermal stability. Besides, elastic constants are calculated, and there are six key parameters: $C_{11}$ = 491.451, $C_{12}$ = 98.396, $C_{13}$ = 0.575, $C_{33}$ = 3.794, $C_{44}$ = 1.871 and $C_{66}$ = ($C_{11}$- $C_{12}$)/2 = 196.528 N/m. According to the Born elastic stability criteria of hexagonal crystal system ($C_{11}$>|$C_{12}$|, $C_{13}^2$<$C_{33}$($C_{11}$+$C_{12}$), $C_{44}$>0 and $C_{66}$>0), GdB$_2$N$_2$ is mechanically stable [38,39].

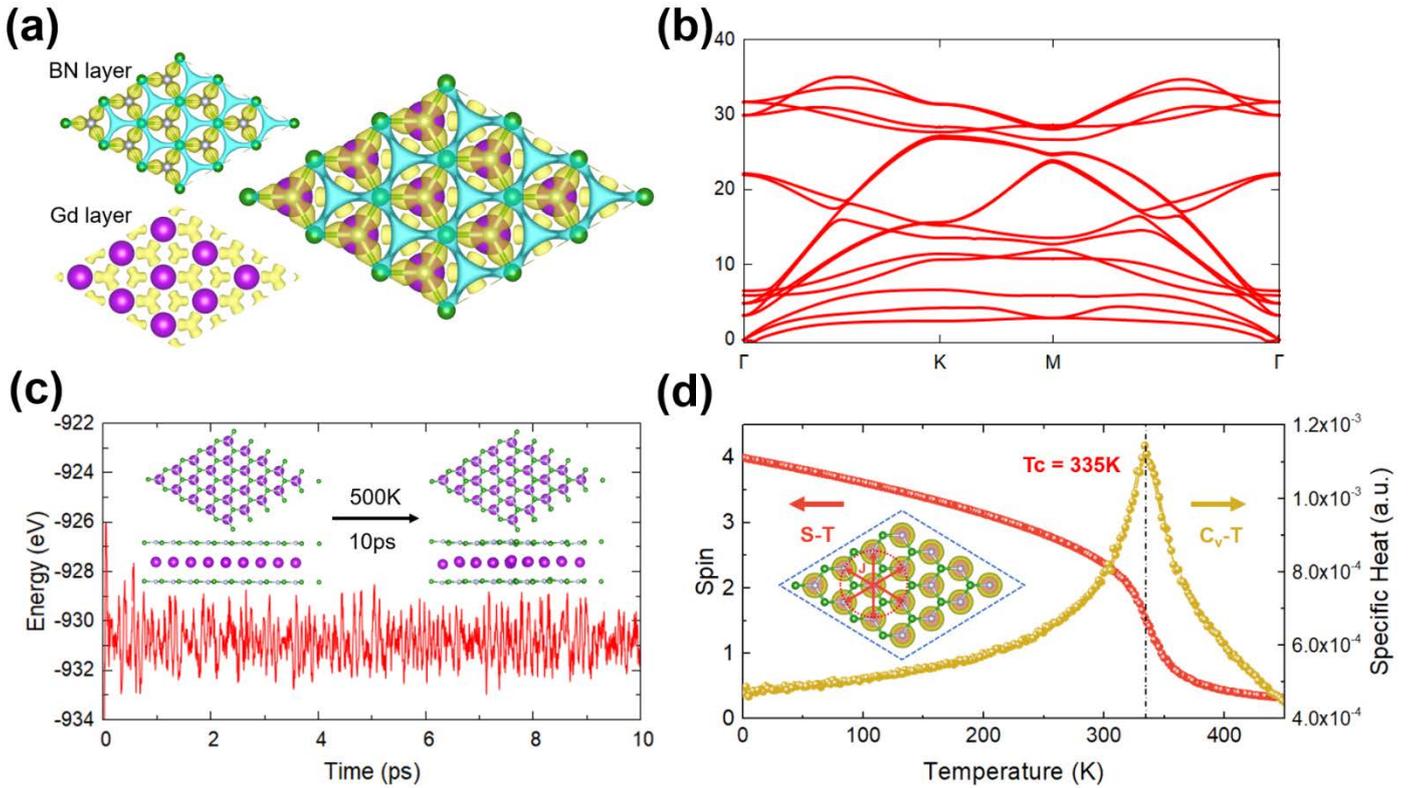

**Figure 2 (a) Electron density difference of the whole GdB$_2$N$_2$ (on the right), BN and Gd layer (on the left), (Yellow regions indicate gaining electrons and cyan regions indicate losing electrons.) (b) phonon spectrum, (c) AIMD and (d) spin and specific heat as functions of temperature. (Spin-dependent electron density is shown in the inner panel.)**

In the above analysis, GdB$_2$N$_2$ has been determined as the intrinsic ferromagnet. The net magnetic moment of each Gd is 7.87 μ$_B$. In addition, spin-dependent electron density is calculated in figure 2(d). It can be found that spin electrons are mainly distributed around Gd ions, which confirms the magnetism mainly attributes to Gd. And due to symmetry of structure, the number of the nearest-neighbor (NN) Gd ions around the central Gd is six. Then, for estimating the Curie temperature of GdB$_2$N$_2$, Monte Carlo (MC) simulation with Metropolis algorithm based on the Heisenberg model is applied to describe thermal dynamics of magnetism in equilibrium states as shown in figure 2(d) [40]. A large 32×32 supercell is construct. 80000 sweeps and 640000 sweeps are used to thermalize the system into equilibrium and obtain all statistical results, respectively. The spin Hamiltonian of the systems is defined as:

$$H = -J \sum_{i,j} S_i \cdot S_j - A(S_i^z)^2$$

where $J$ is the nearest neighbor exchange parameter, $S_i$ and $S_j$ are the spin operator, $S_i^z$ is the spin component parallel to the z direction, and A is the anisotropy energy parameter. Besides, the specific heat capacity C$_v$ is defined as $C_v = \frac{(\langle E^2 \rangle - \langle E \rangle^2)}{K_b T^2}$. The magnetic exchange parameter $J$ is calculated to be 1.547 meV. According to spin and specific heat capacity curve plotted in figure 2(d), finally the Curie temperature is calculated to be 335 K. With the above same method, the Curie temperature of GdI$_2$ has been calculated to be 231 K as shown in figure S2, similar to the previous work of 241 K and 251 K [22,23]. So the correctness of this method can be verified.

Exploring the influence of external stimulus on magnetic properties of materials can broaden the practical applications and play an important role in adapting to the complex situations, e.g. lattice mismatch. As shown in Figure 3(a), biaxial strains ranging from -0.5% to 5% are applied on GdB$_2$N$_2$ to explore the variation of $\triangle$E (energy difference between the FM and AFM1 states) and MAE. The $\triangle$E monotonously increases under compressive strains and attains 221.22 meV/f.u. at the -0.5% compression, while decreases upon tension and drops to 88.09 meV/f.u. at the 5% tension. The nearly linear curve shows the good modulation effect of external strains on $\triangle$E of GdB$_2$N$_2$. While the variation of MAE under different strains is more complex. The MAE has the highest value of 10.38 meV/f.u. under no strain, and reaches the lowest value of 8.04 meV/f.u. under 3% tensile strain. The larger strain (up to 40% tensile strain) has also been applied to affect $\triangle$E of GdB$_2$N$_2$ as shown in figure S3(a). When the biaxial strain reaches 20%, the value of $\triangle$E becomes negative; while when the strain reaches up to 35%, $\triangle$E returns to be positive again. In figure S3(b), the estimated magnetic exchange parameter J also fluctuates with strains between FM and AFM coupling, which is consistent with an RKKY-type interaction. Besides, the influence of charge-carrier doping ($\leq$ 0.5e/h per f.u.) on $\triangle$E and MAE is discussed and plotted in figure 3(b). It can be found that the hole doping plays a positive role in promoting $\triangle$E, while the electron doping plays a negative role. Whether with electron or hole doping, the value of MAE monotonously decreases, except the role of hole doping shows more significant. The influence of biaxial strains and charge-carrier doping on magnetic moment of GdB$_2$N$_2$ has also been analyzed in figure S4. Tensile strains and charge-carrier doping both can make magnetism monotonously decrease. And the role of hole doping shows more dominant than electron doping.

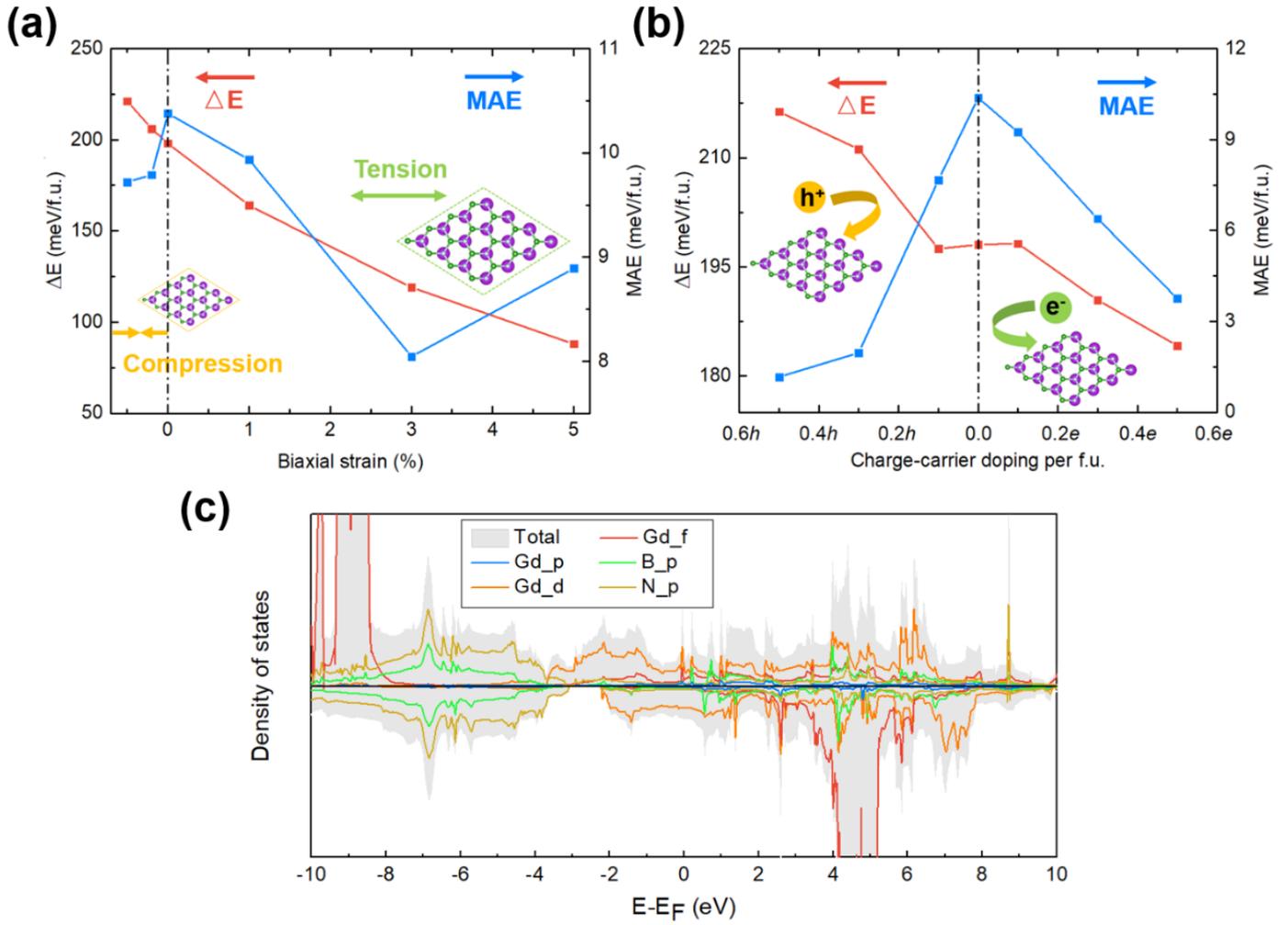

**Figure 3 (a) △E (△E = $E_{AFM1}$ – $E_{FM}$) and MAE (MAE = $E_{soc(z)}$ – $E_{soc(in-plane)}$) as functions of biaxial strains, (b) △E and MAE as functions of charge-carrier doping concentration and (c) projected density of states.**

Then, the electronic properties of $GdB_2N_2$ are analyzed. The calculated projected density of sates is shown in figure 3(c). (The contributions of each element and their s orbital to density of states are shown in figure S5(a) and (b), respectively.) The localized 4f states of Gd is away from the Fermi level and mainly distributed in the region between 3 eV and 6 eV, or between -10 eV and -8 eV. And they do not mix noticeably with the valence-band states. Thus, the 4f states are indeed strongly localized and can only interact via an RKKY-type of interaction. The DOS at $E_F$ mainly attributes to d states of Gd, which leads to metallicity of $GdB_2N_2$ and also show a significant spin polarization. This is also in line with RKKY-type magnetic interactions. For further research, the band structures of $GdB_2N_2$ have been calculated, projected band structures without and with considering SOC effect are plotted in figure 4(a) and 4(b), respectively. It can be found more clearly that Gd element mainly contributes to the metallicity of $GdB_2N_2$. The interactions between Gd and B ions are strong, and parts of electrons in B also exist around Fermi level. In contrast to band structures of semiconductor *h*-BN [33], Gd element plays a role as donors to provide extra electrons, shift the Fermi energy to higher value and also interacts with B ions in BN layer. Moreover, there are several Dirac points below and above Fermi surface, five of which are marked with blue dotted circles. They have an n-type or p-type linear dispersion in figure 4(a), but can show a small gap after considering SOC effect in figure 4(b).

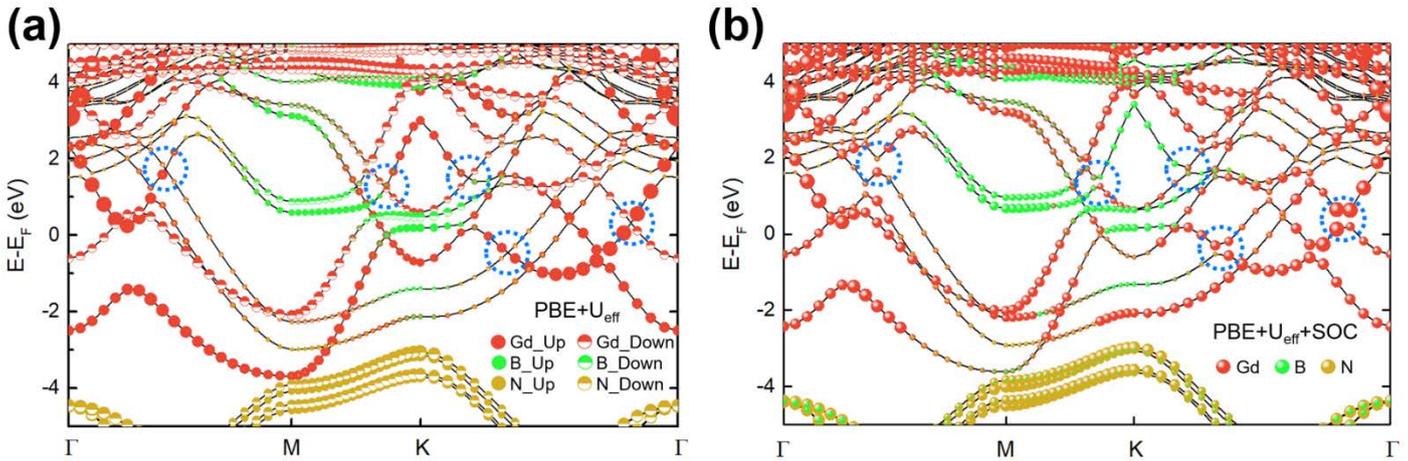

**Figure 4. Projected band structures (a) without SOC effect and (b) with SOC effect. Blue dotted circles indicate the Dirac points.**

Finally, we propose a possible route to experimentally synthesize the present predicted $GdB_2N_2$. As has been introduced, $GdB_2N_2$ consists of three layers, and the Gd layer is sandwiched by two BN layers. In the previous works, monolayer BN has already been found with excellent stability and mature synthetic methods, so BN nanosheet is proper as ideal substrate [14]. The possible synthesis process is as follows: firstly, adsorbing Gd atoms upon

BN monolayer can be achieved by means of molecular beam epitaxy (MBE) method; then dry transfer method is used for another BN monolayer to cover the previous Gd/BN heterojunctions. The three possible Gd adsorption sites are marked in figure S5(a). The adsorption energy is defined as: $E_{adsorption} = E_{GdBN} - E_{Gd} - E_{BN}$, where $E_{GdBN}$, $E_{Gd}$ and $E_{BN}$ represent the total energy of Gd adsorbed $h$-BN nanosheet, single Gd atom and monolayer $h$-BN. Because adsorption of Gd at site I is more energetically favorable than other sites as shown in figure S5(b), we can conclude that Gd ions prefer to be adsorbed at site I. Therefore, the 2D $GdB_2N_2$ can be experimentally realized by adsorbing Gd on one BN monolayer and then covering another BN monolayer.

## 4. Conclusions

In conclusion, we have predicted a novel 2D rare-earth material – $GdB_2N_2$ based on DFT calculations. 2D $GdB_2N_2$ exhibits metallicity, room-temperature ferromagnetism and excellent stability. Monte Carlo simulations manifest that the Curie temperature of the 2D $GdB_2N_2$ material is as high as 335 K, showing the promising applications in spintronics devices, and the MAE is demonstrated to be 10.38 meV/Gd. Ab initio molecular dynamics (AIMD), phonon spectrum and elastic constant matrix are also calculated to demonstrate the thermal and mechanical stability of the 2D $GdB_2N_2$ respectively. The $GdB_2N_2$ can maintain the initial structure even at 500 K after 10 ps. The external stimulus including biaxial strains and charge-carrier doping can tune the magnetic properties of $GdB_2N_2$. The ferromagnetism is finally found to originate from the RKKY interactions of Gd atoms. The discovery of 2D ferromagnetic $GdB_2N_2$ here paves the way for development of modern spintronics and high-performance spintronic devices.

**Credit author statement**

HY Tan performed the calculations, summarized the results, and wrote the draft. GC. Shan, and JL. Zhang discussed the data. GC. Shan designed the experiments, revised the manuscript, and supervised the entire study. All authors checked the manuscript.

**Declaration of competing interest**

The authors declare that they have no known competing financial interests or personal relationships that could have appeared to influence the work reported in this paper.

**Acknowledgments**

This work was supported by the National Natural Science Foundation of China (51971057) and the Research Funds for the Central University.